# On Spatial Lag Models estimated using crowdsourcing, web-scraping or other unconventionally collected data

Arbia, G.[1] and Nardelli, V.[2]

[1] Catholic University of the Sacred Heart (Milan – Rome, IT)
[2] University of Milan-Bicocca (Milan, IT)

***Abstract***: *The Big Data revolution is challenging the state-of-the-art statistical and econometric techniques not only for the computational burden connected with the high volume and speed which data are generated, but even more for the variety of sources through which data are collected (Arbia, 2021). This paper concentrates specifically on this last aspect. Common examples of non traditional Big Data sources are represented by crowdsourcing (data voluntarily collected by individuals) and web scraping (data extracted from websites and reshaped in a structured dataset). A common characteristic to these unconventional data collections is the lack of any precise statistical sample design, a situation described in statistics as "convenience sampling". As it is well known, in these conditions no probabilistic inference is possible. To overcome this problem, Arbia et al. (2018) proposed the use of a special form of post-stratification (termed "post-sampling"), with which data are manipulated prior their use in an inferential context. In this paper we generalize this approach using the same idea to estimate a Spatial Lag Model (SLM). We start showing through a Monte Carlo study that using data collected without a proper design, parameters' estimates can be biased. Secondly, we propose a post sampling strategy to tackle this problem. We show that the proposed strategy indeed achieves a bias-reduction, but at the price of a concomitant increase in the variance of the estimators. We thus suggest an MSE-correction operational strategy. The paper also contains a formal derivation of the increase in variance implied by the post-sampling procedure and concludes with an empirical application of the method in the estimation of a hedonic price model in the city of Milan using web scraped data.*

---

[1] Corresponding author. Email address: `giuseppe.arbia@unicatt.it`

# 1. Introduction

Until only few years ago researchers in spatial econometrics were mostly interested in estimating models based only on regional data. However, as it is well known, inference based on regional aggregates is intrinsically biased by the presence of the Modifiable Areal Unit Problem (or MAUP), in that results are influenced by both the geographical scale of analysis and the geographical grouping criterion adopted (see Gehlke and Bihel, 1934; Arbia, 1989 and Arbia and Petrarca, 2011). More recently, the increasing availability of individual geo-coded micro-data has shown a way to tackle the MAUP giving rise to the new branch termed *spatial microeconometrics* (Arbia et al, 2021). A big impulse in this direction has been given by the advent of the Big Data revolution and the contemporaneous availability of previously unconceivably rich datasets to test the theoretical models. Following the celebrated definition of Laney (2001), Big Data are characterized by three V's, namely: Volume, Velocity and Variety (Arbia, 2021). While the characteristics of Volume and Velocity pose serious challenges to statistical techniques from a computational point of view when dataset are very large and collected in almost real-time, the third *V* (Variety) creates other kinds of problems, in terms of the data generating mechanisms, which require formal solutions prior the application of the standard econometric techniques.

It should be emphasized right at the beginning of this paper, that our interest is restricted to the problems raised by the increasing use of a large variety of unconventionally collected datasets in spatial data analysis and that, therefore, the examples we have in mind are not necessarily related to very large and quickly growing databases. Our aim is to shed light on the distortions which derive from the unwise use of data of different varieties in spatial econometric modelling and to suggest possible solutions. Consequently, we will place no concern on the computational issues that may arise from the application of our proposed methodology to very large quantity of data.

When considering new unconventional data sources, two very common examples are represented by *crowdsourcing* (data voluntarily collected by individuals) and *web scraping* (data extracted from websites and reshaped in a structured dataset). These typologies of data with the addition of a spatial reference are also known as Volunteered Geographic Information (VGI) (Goodchild, 2007). Crowdsourced data are common in many situations. An example is represented by data collected through smart phones, in order to measure phenomena that are otherwise difficult to quantify precisely and timely. Among these, e. g., the crowdsourced collection of food prices in developing countries (see, e.g. Arbia et al. 2018) and epidemiological data (Crequit et al., 2018). The practice of extracting data from the web and using them in statistical analyses is also becoming more and more popular, e. g. to collect online prices in the real estate market (Beręsewicz, 2015; Boeing et al., 2017) or in consumer goods (Cavallo et al, 2016).

A common characteristic of these two unconventional data collection sources is represented by the lack of any precise statistical sample design. In crowdsourcing the participation is generally voluntary thus producing a self-selection of the collectors. A similar problem emerges when extracting data that were published on web platform and social media without taking into account the process that led to their publication.

This situation is described in statistics as "convenience sampling" in the presence of which, as it is well known, no sound probabilistic inference is possible (Hansen et al, 1953). More precisely, while in a formal sample design the choice of observations is suggested by a precise mechanism which allows the calculation of the probabilities of inclusion of each unit (and, hence, sound probabilistic inference), on the contrary with a *convenience* collection no probability of inclusion can be calculated thus giving rise to over- under-representativeness of the sample units.

When estimating a population mean, Arbia et al. (2018) proposed a technique termed "*post-sampling*", a special form of post-stratification (Holt and Smith, 1979; Little, 1993) in which data are reweighted prior their use in an inferential context. In this paper we aim to generalize such an approach using the idea of post-sampling when estimating a Spatial Lag Model (SLM, see Arbia, 2014).

The layout of the paper is the following. In Section 2 we introduce more formally the idea of post-sampling in the general context of parameter estimation. Section 3 is devoted to show the implication of using data gathered with convenience sampling when estimating a spatial econometric model. In particular, we show through a Monte Carlo study, that by using data collected without a proper design, the parameter estimates in a SLM become biased and that the bias can be reduced by introducing an appropriate post-sampling procedure though at the cost of increasing the estimator's variance. In Section 4 we will present an operational version of post-sampling leading to an MSE-reduction mechanism for the case of the Spatial Lag Model. This mechanism is then used in Section 5 in an empirical application which aims at estimating an hedonic price model using data that were scraped from the web visiting the real estate company advertisements in Milan (Italy). In the empirical example, we show that it is possible to fine-tune the post-sampling so as to find a compromise between the bias and the increase in the variance of the estimators. The paper concludes with some final comments contained in Section 6 and with an Appendix reporting the formal derivation of the increase in the estimation variance which is implied by the post-sampling procedure.

## 2. The post-sampling MSE-correction procedure

It can be traced back to the early contributions of Sir Ronald Fisher (Fisher, 1935) the idea that if we aim at a satisfactory generalization of the sample results, the sample experiment needs to be rigorously programmed. Indeed, lacking a formal sample design, data have to be considered *convenience* (or sometimes *judgment*) sampling with which no probabilistic inference is possible (Hansen et al., 1953). This problem emerges dramatically in the Big Data era when we increasingly avail huge quantities of data which, almost invariably, do not satisfy the necessary conditions for probabilistic inference. In recent years researchers are becoming aware of this problem trying to suggest solutions to reduce the distorting effects inherent to non-probabilistic designs (Fricker and Schonlau, 2002). One possible strategy that has been suggested is to discard some of the redundant data and to concentrate only on smaller datasets obtained as a rigorously programmed sample. This strategy can be drawn back to the ideas expressed by the philosopher of science Jules-Henri Poincare' long before the advent of Big Data: "*If the scientist had at his disposal infinite time, it would only be necessary to say to him: 'Look and notice well'; but, as there is not time to see everything, and as it is better not to see than to see wrongly, it is necessary for him to make a choice*" (Poincare', 1908). Following this basic idea that "it is better not to see that see wrongly", Arbia *et al.* (2018) suggested to transform crowdsourced datasets by discarding observations in such a way that they resemble a formal sample design. This procedure has been termed *post-sampling*.

There are three forms of post-sampling. The first one consists in drawing randomly a subset of the units from the original dataset obeying some formal sampling design (*hard-core* post sampling). In this way, however, in discarding part of the data, while correcting for the bias, the procedure is doomed to produce an obvious increase in the variance of the estimators due to the diminished sample size. To moderate this effect a second form of post-sampling consists in discarding only some of the data from the original dataset in order to approximate a formal sample design, but without fully achieving the desired configuration (*flexible* post sampling). Finally, in a third form of post-sampling no observation is discarded, but all the original data are re-weighted taking into account the requirements of a formal sampling design (*weighted* post sampling, Arbia et al. 2018).

In this section we will briefly review the three strategies.

In the nutshell the *hard-core* post sampling method can be described as follows. Suppose that, say, $n_l$ ($l = 1, \dots L$) observations are collected with a convenience sampling on a set of *L* given geographical locations (e. g. counties or regions) in such a way that the total number of observations available in the whole area is $n = \sum_{l=1}^{L} n_l$. To implement the post-sampling strategy, we then compare the location of the observed data with that of a set of points required if a formal sample design of equivalent sample size was applied. While in principle any formal design can be adopted in this phase, an obvious choice is to consider a geographically random stratified design with probability proportional to size weights (PPS) using some convenient stratification variables. In many geographical studies observations may be stratified considering some auxiliary variable such as, e. g., habitat type, elevation, population density, soil type or other. Human populations may be stratified on the basis of variables like city size, sex, or other socioeconomic factors. (Thompson, 2012).

Let us call $m_l$ the number of observations which are required by the chosen sample design in each geographical location with $\sum_{l=1}^{L} m_l = n$. In each of the *L* geographical locations considered, some data are then deleted at random so that the number of observations available after this post-sampling operation, say $\bar{n}_l$, will be proportional to the desired sample, so that $\bar{n}_l = k\, m_l$, $k \in [0,1]$ being a proportionality constant. In this way the transformed dataset corresponds to a geographically stratified sample for which it is possible to calculate the probabilities of inclusion and, hence, to draw probabilistic inference. However, the price to pay for the bias reduction is a loss of $(1-k)\,\%$ of the originally available observations which will in turn produce an obvious increase in the variance of the estimators.

The *hard-core* post-sampling procedure is too rigid and a more flexible version can be implemented allowing only a partial reduction of the number of units thus compromising between the reduction of bias and the increase in the estimator's variance. To this aim, the number of observations deleted in each location can be limited by a parameter, say $\zeta$, which measures the degrees of discrepancy between the post-sampled dataset and the dataset required by the formal sample design chosen as a benchmark.

In particular, when $\zeta = 0$ the final dataset will exactly be equal to the original one and no post-sampling correction is considered. Conversely, when $\zeta = 1$ the flexible post-sampling will coincide with its *hard-core* version. Varying the value of the post-sampling parameter $\zeta$ in the range between 0 and 1 it is possible to create different datasets compromising between bias and estimation variance (Nardelli, 2019).

To tackle the problem deriving from the increasing estimation variance, a further post-sampling strategy can be proposed which considers a re-weighting of the available dataset.

To implement this strategy, we can calculate, in each geographical location, a *post-sampling ratio (PS)*, defined as the ratio between the number of observations available required by the reference sample design through the convenience sampling and those collected with a convenience criterion, say $PS_l = m_l/n_l$. The estimation of a population parameter is then obtained by considering a weighted version of the dataset using the post-sampling ratio as weights. Thus, if $PS_l$ = 1 the number of observations available in each location *l* are exactly those required by the sampling design and no adjustment is needed. The available observations will instead have to be over-weighted If $PS_l > 1$ and, on the contrary, down-weighted when $PS_l < 1$. This strategy has been adopted in Arbia et al. (2018) in order to estimate the food price index in Nigeria using crowdsourced data. It can be argued that the hard-core and the flexible approach can be nested into the weighted approach. Indeed, the only difference between the three approaches is that in the hard-core and flexible case the weights can only be either 0 or 1 and weights are assigned randomly whereas in the weighted approach the applied weighting allows for deviations from the proportionality to the ideal sampling design.

In the next section we will illustrate the use of a flexible post-sampling procedure in order to estimate a spatial lag model whereas we leave to a future work the development of the theory behind the weighted post-sampling method in spatial econometric modelling.

**3. Effects of post-sampling in the estimation of SLM parameters**

In this section we report the results of a set of Monte Carlo experiments designed to study what are the effects of using convenience spatial data (such as web-scraped or crowdsourced data) in order to estimate the parameters of a spatial econometric model.

In our simulations we considered, in particular, the following Spatial Lag Model (SLM, see Arbia, 2014):

$$y = \beta x + \rho W y + \sigma^2 \varepsilon \qquad (1)$$

where y is an n-by-1 vector of observations of the dependent variable y, x is an n-by-1 vector of observations of the independent variable (for the sake of simplicity we consider only one predictor in our model), $\varepsilon \sim i.i.d. N(0,1)$ are the independent innovations and $\beta, \rho$ and $\sigma^2$ are scalar parameters. In Equation (1) W is an exogenously specified weight matrix taking care of the links of proximity between the n units. In particular, the weights are derived using the distance-threshold neighboring criterion (Arbia, 2014), by selecting a threshold that guarantees the absence of isolated points.

In a first Monte Carlo experiment, we considered two different situations where we have a population of $N = 5600$ points (Simulation 1) and $N = 11200$ (Simulation 2) randomly distributed with different densities in the four quadrants of a unitary square. We further assumed that, in both instances, only n<N of such observations are available following a convenience sampling. In all our experiments we set the sample size n = 270 (See Table 1 and Figure 1) thus considering two cases of different sample proportion which is equal to 4.8 % and 2.4 % respectively.

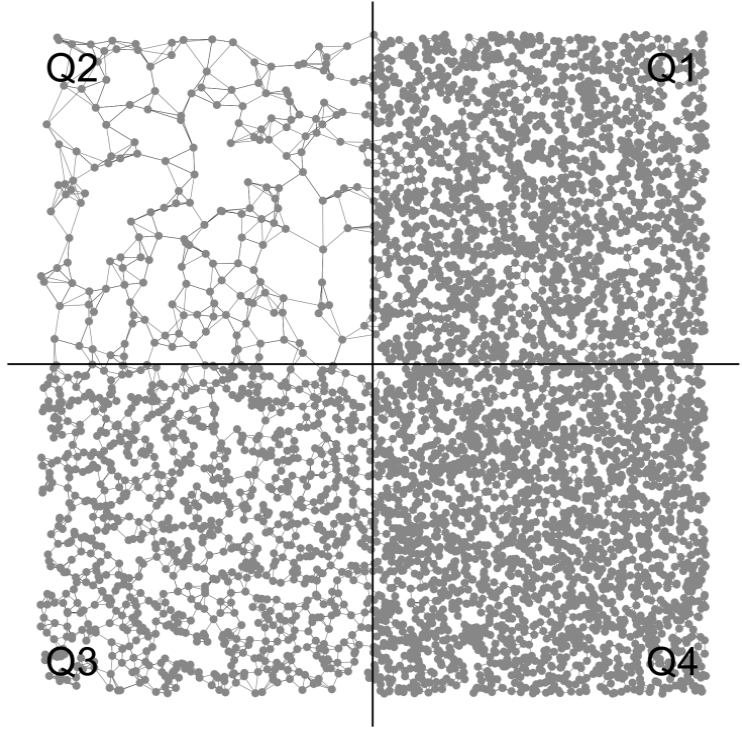

Figure 1: Spatial distribution of N= 5600 points randomly distributed in the four quadrants of a unitary square and their links of neighborhood.

|  | Q1 | Q2 | Q3 | Q4 |
|---|---|---|---|---|
| N (Sim 1) | 2000 | 200 | 1000 | 2400 |
| N (Sim 2) | 4000 | 400 | 2000 | 4800 |
| n (Sim 1-2) | 70 | 20 | 150 | 30 |
| Sample proportion (Sim 1) | 3.50% | 10.00% | 15.00% | 1.25% |
| Sample proportion (Sim 2) | 1.75% | 5.00% | 7.50% | 0.625% |

Table 1: Population, sample size and sample proportion in the four quadrants reported in Figure 1.

We then simulated the vector of observations of the exogenous regressor from a Gaussian distribution with expected value = 10 and variance = 1. This vector is kept constant in each simulation run, whereas, at each run, the error term $\varepsilon$ is generated from a standardized normal distribution. In each replication of the Monte Carlo experiments, the vector y is then generated through the following expression:

$$y = (I - \rho W)^{-1}\beta x + \sigma^2 (I - \rho W)^{-1}\varepsilon \qquad (2)$$

In each replication, a *flexible post-sampling* procedure is performed and the resulting dataset is used to estimate the Spatial Lag Model. In the Monte Carlo experiments the simulation parameters $\rho$ and $\zeta$ can assume the following values: $\rho = [0; 0.2; 0.4; 0.6; 0.8]$ and $\zeta = [0; 0.2; 0.4; 0.6; 0.8; 1]$. The parameter $\beta$ is constant in all simulations.

It is not trivial to being able to simulate the process of convenience sampling given the lack of regularity which is implied in the data collection mechanism. In this paper, in order to mimic the process of a convenience sampling we considered a marked departure from the geographically random stratified case where webscraped or crowdsourced data are less frequent in the more densely populated quadrants and comparatively more frequent in the low-density quadrants. In particular, for Simulation 1, we considered the following sampling proportions in the four quadrats: Q1 = 3.5 % of units, Q2 = $10\%\ units$, Q3 = 15% and Q4 = 1,25%. For Simulation 2, the proportions become: Q1 = 1.75 % of units, Q2 = $5\%\ units$, Q3 = 7.5% and Q4 = 0,625%. Our process mimics, e. g. an imitation effect among the voluntary collectors in a crowdsourcing experience, as it is observed, e. g., in Arbia et al. (2018), or simply a selection bias linked to some geographical feature of the webscraping process.

The aim of our Monte Carlo experiment is to monitor the sensitivity of the performances of the various estimators to the post-sampling parameter $\zeta$ (see Section 2) both in terms of bias and of the estimation variance. To this aim, first of all, we estimated Model (1) through Maximum Likelihood using all available observations without any post-sampling correction ($\zeta = 0$). We then introduced progressively a post-sampling correction mechanism using various levels of the post-sampling parameter $\zeta = (0.2; 0.4; 0.6; 0.8; 1)$. As already stated, the case of $\zeta = 1$ corresponds to the *hard-core* post-sampling.

Table 2 reports the average number of observations that survive the post-sampling procedure for each level of the parameter $\zeta$.

Table 2: Sample size as a function of parameter $\zeta$ in the Monte Carlo experiment.

| $\zeta$ | Sample size |
|---|---|
| 0 | 270 |
| 0.2 | 222 |
| 0.4 | 174 |
| 0.6 | 126 |
| 0.8 | 89 |
| 1 | 70 |

With regards to the estimation of the parameter $\beta$, Figure 2 displays the average behavior of the bias, of the estimation variance and of the MSE implied by each level of post-sampling at the various levels of the spatial correlation parameter $\rho$ in 500 simulation replications in each of the two simulations.

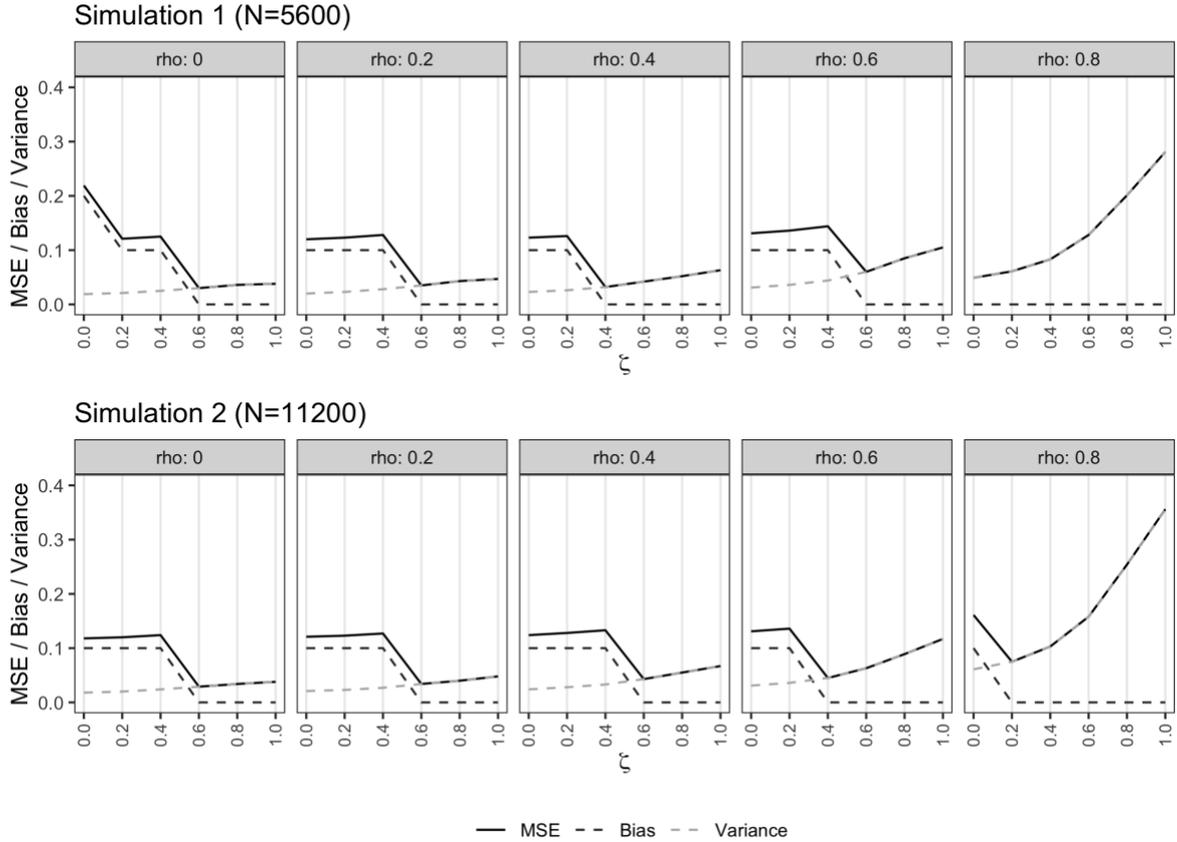

Figure 2: Simulation results: Squared bias, estimation variance and MSE of the estimators of $\beta$ in a SLM for the two simulation cases, for different levels of the parameters $\rho$ and $\zeta$.

Figure 2 clearly reveals that, without any post sampling correction (case of $\zeta = 0$), the ML estimates of the parameter $\beta$ in Equation (1) present a marked upward bias which seems to be rather independent of the level of the spatial correlation parameter $\rho$ (only with a peak at $\rho = 0$ in Simulation 1. In this situation the estimation process requires a post-sampling correction in order to restore the condition of unbiasedness that went lost with the convenience sampling. Inspecting Figure 2 we also notice that the (squared) bias decreases monotonically to 0 when $\zeta$ increases towards 1. This behavior shows that the proposed procedure is effective in achieving the desired bias-correction mechanism. However, Figure 2 also shows that the procedure produces progressively a higher variance of the estimators which is intuitively due to the reduced number of observations available. Furthermore, Figure 2 shows that, in each simulation case, the MSE criterion identifies an optimal level of $\zeta$ which compromises between bias and estimation variance. In our experiments this value is at a level $\zeta = 0.6$ ($\zeta = 0.2/0.4$ when $\rho$ is very high).

Finally, Figure 2 allows a comparison in the performances of the estimation procedure proposed in the two simulation cases (Simulation 1 and 2) where we considered two different sample proportions. Indeed, by comparing the results of Simulation 1 with those of Simulation 2, we observe that the post-sampling procedure seems to be unaffected by the sample proportion. The only differences we observe in the graphs are when $\zeta = 0$, that is when no post-sampling correction is applied.

In order to provide some insights about the impact of various specifications of the W matrix on the properties of the estimators, in a second simulation exercise, we repeated our Monte Carlo experiment using different definitions of the spatial weight matrices. In particular, we adopted again a threshold distance (TR) definition as in the first exercise, and we compare the results using a 4 k-nearest neighbours (KNN) and an inverse distance (ID) definition as possible alternatives. With this choice we also are able to provide a comparison between a low density W matrix (the 4 nearest neighbours has a density of 8/n), a high density (the inverse distance definition is associated with a density of n/(n-1)) using the case of the threshold distance definition as an intermediate case .

The results obtained in this second Monte Carlo experiment are summarized in Figure 3. Our experiments show that the variance of the estimators is almost unaffected by the choice of the W matrix with only a slightly larger estimation variance observed when we employ an inverse distance definition and when the post-sampling is introduced with a low $\zeta$ ($\zeta <0.4$). In contrast, the bias of the estimator is affected by the choice of W. Indeed, again only for low values of $\zeta$ ($\zeta =0.2$), we observe a higher bias for the knn definition and a lower bias for the inverse-distance definition with the threshold-distance case in an intermediate position. In this respect there seems to be a relationship between bias and the sparsity of W with a higher bias observed in sparser configurations. When the post-sampling correction is stronger, however ($\zeta \geq 0.4$), all three definitions lead approximately to the same results both in terms of the bias and of the variance of the estimators. Figure 3 also reveals that the minimum MSE is achieved at $\zeta \geq 0.6$ using the threshold distance or the knn criterion, whereas it remains constant in the interval between 0.2 and 0.6 if we employ the inverse-distance specification of W.

In summary, the inverse-distance definition produces the highest increase in the estimation variance, but also an immediate reduction of the bias as soon as a post-sampling correction is introduced. In terms of the MSE, inverse-distance connectivity matrices have to preferred at low values of $\zeta<0.6$, while for higher values the other two specifications guarantee a lower MSE.

It should be noted that, in comparing the result obtained with different W matrices, a problem of miss-specification may emerge in the process of post-sampling. Indeed, when data are discarded in the post-sampling process, the W matrix needs to be respecified and it could happen that data generated by a certain, say, $W_1$ matrix are eventually modeled by, say, a $W_2$ matrix simply because the reduction of the number of observations is too large. In this respect the knn specification seems to guarantee more robustness at high values of $\zeta$, because its structure remains always the same (both in terms of its density and of its single entries) regardless the number of observations.

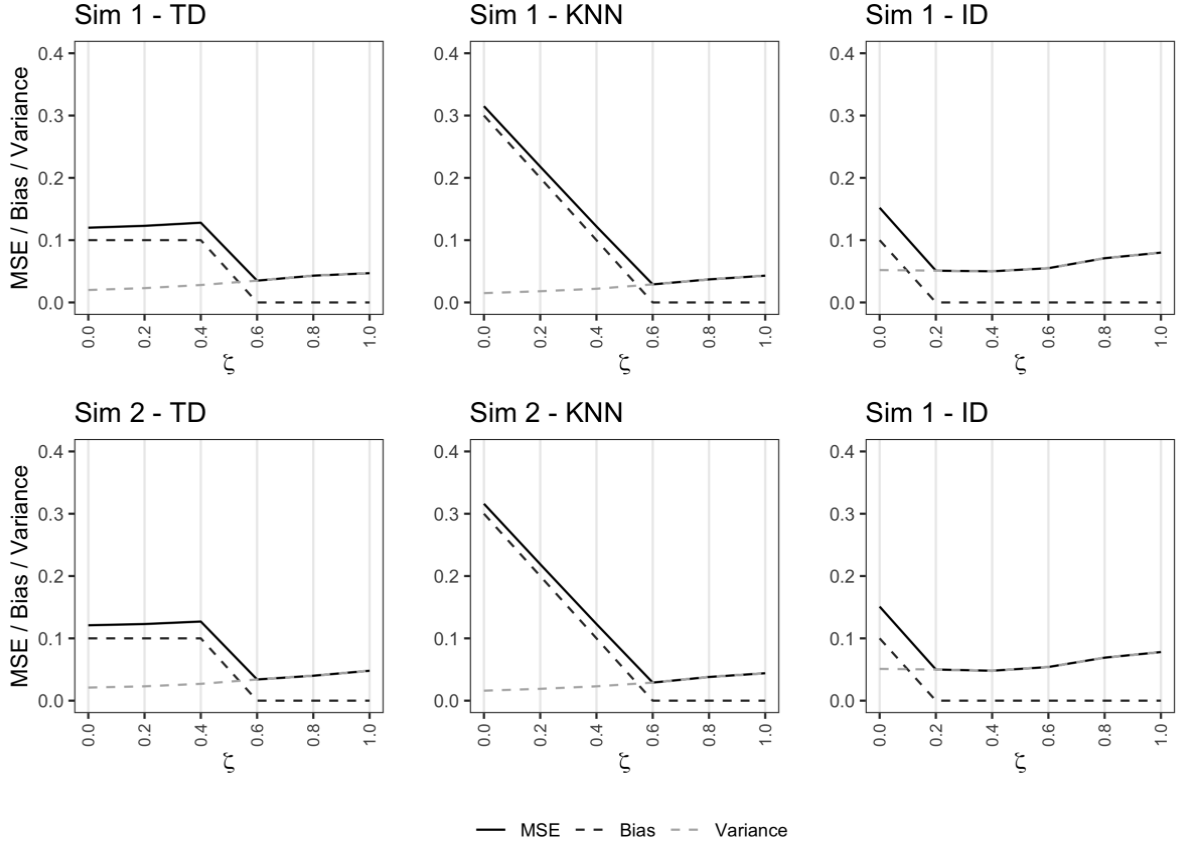

Figure 3: Comparison between three different W matrix specifications of Bias, Estimation variance and MSE. TD = threshold distance; KNN= 4 k-nearest neighbours; ID = inverse-distance. In all cases $\rho = 0.2$.

**4. An operational procedure to apply the post sampling MSE-correction mechanism**

Although in the previous section we considered quite general simulation cases, in principle, we might expect that in different empirical situations the level of $\zeta$ which optimally compromises between bias and sampling variance could be different and this should be evaluated prior the process of estimation.

In order to identify such an optimal value in practical cases we can adopt the following operational procedure.

First of all, let us call $\hat{\beta}_\zeta$ the MLE of the parameter $\beta$ obtained by making use of a *flexible post-sampling* correction mechanism with a post-sampling parameter $\zeta$. Secondly, from the simulations reported in the previous section, we have noticed the tendency of the estimation bias towards 0 when $\zeta$ goes to 1 (see Figure 2). For this reason, we can assume that the bias is minimized when $\zeta = 1$ and that a good estimation of the bias in the estimation of the parameter $\beta$ can be provided by $B(\hat{\beta}, \zeta) = |\hat{\beta}_\zeta - \hat{\beta}_1|$ .

Once the bias is estimated, in order to calculate the expression of the MSE to be minimized, we also need an estimation of the standard error of the estimator $\hat{\beta}$.

In this respect, if we consider the ML estimator of the paramteres of the Spatial Lag Model reported in Equation (1), we can rely on an asymptotic result. Indeed, in this case, the Hessian associated to the likelihood function has been formally derived (see Anselin, 1988; Lee, 2004).

By using such result, in any specific empirical situation, we can evaluate the Fisher information matrix and, through this, the expected asymptotic variance of the estimator of $\beta$ for any given value of $\zeta$, distinctively for the model estimated on the original dataset and for the one based on its post-sampled version. Let us call these two asymptotical variances $AVar(\hat{\beta})$ and $AVar(\hat{\beta})_\zeta$ respectively, with an obvious symbolism. In the case of a Spatial Lag Model these asymptotic results are reported in the Appendix to the paper.

Making use of these foregoing results, we can identify the optimal level of post sampling by employing the following operational steps.

STEP 1. First of all, we estimate the regression parameter $\beta$ of the Spatial Lag Model in Equation (1), using MLE at any given value of $\zeta$. Let us refer to this set of estimates as $\hat{\beta}_\zeta, \zeta \in [0; 1]$.

STEP 2. Secondly, we estimate the bias associated to each level of the post-sampling parameter $\zeta$ as $B(\hat{\beta}, \zeta) = |\hat{\beta}_\zeta - \hat{\beta}_1|$ assuming that for $\zeta = 1$ the upward bias due to the convenience collection is minimized.

STEP 3. Thirdly, for each level of $\zeta$, we estimate the asymptotic variance of $\hat{\beta}$ using Equation (A8) in the Appendix. We call it $AVar(\hat{\beta})_\zeta$

STEP 4. Fourthly, we estimate the MSE of the estimator of $\beta$ for each level of the post-sampling parameter $\zeta$ by using the expression:

$$MSE(\hat{\beta}, \zeta) = (\hat{\beta}_\zeta - \hat{\beta}_1)^2 + AVar(\hat{\beta})_\zeta \qquad (3)$$

STEP 5. We then identify the optimal level of the post-sampling parameter $\zeta$ (say $\tilde{\zeta}$) through the expression:

$$\tilde{\zeta} = argmin_{\zeta \in [0,1]}\{MSE(\beta, \zeta)\} \qquad (4)$$

STEP 6. Finally, we pick the optimal $\zeta$ as our final estimate and we estimate the parameter $\beta$ using the MLE corresponding to that particular value of $\zeta$, say $\hat{\beta}_{\tilde{\zeta}}$.

The practical use of the procedure described in this section will be illustrated in the next section making use of some real house price data scraped from the web.

**5. An empirical application: estimating a spatial hedonic price model with webscraped data**

In the Monte Carlo study reported in Section 3, the optimal value of $\zeta$ to be used in the post-sampling procedure, is obtained with a computational search of the value that minimizes the MSE of the estimators. In this section we aim at showing how the procedure described in Section 4 to identify the MSE-optimal value of $\zeta$ can be made operational when applied to a real set of data.

In particular, a flexible post-sampling strategy will be employed in order to estimate a hedonic price Spatial Lag Model. We will estimate the model using real data that were scraped visiting the real estate company advertisements appearing on the web in the city of Milan (Italy) in May 2019. As it is known, the theoretical foundations of the hedonic price are based on the hedonic utility in theory of consumer's demand. According to Lancaster (1966), the composite goods can be considered as homogeneous, so that the utilities are entirely based on the characteristics of goods. The price then is given by the unit cost of a series of characteristics which belong to two different categories: residential structure (such as its style, lot size, and the number of rooms) and externalities associated with the geographic location (i.e. location effects) (Xiao, 2017). Providing empirical support to the use of spatial econometric models, Can (1992) found that houses that are close in geographic space are likely to have similar attributes. Indeed, the existence of geographical submarkets leads to the definition of spatial models which are able to account for the spatial dependence among the observations. One further reason why house prices may be spatially autocorrelated is that property values in the same neighborhood capitalize shared location amenities.

To gather our working dataset, we performed a systematic web-scraping of all real estate companies in Milan for the whole month of May 2019. The operation of extracting the relevant information from the internet is not an easy task due to the fact that most of the data available on the websites are presented in an unstructured format. The complexity is also related to factors that depend on how the website was designed, on how it works and on the particular technologies used. In particular, in order to analyze the supply side prices in the real estate market in Milan, we used a Python web-scraping algorithm to save the source code of the web page into the HTML format, to extract the useful information parsed through the XPath language and, finally, to store the structured dataset thus obtained after each round of scrape in a relational database management system (RDMS). The rest of the analysis is performed using the commands included in the `R` package `spdep`.

After the systematic webscraping of all real estate companies in Milan for the whole month of May 2019, our final dataset is composed by 1,000 points data observations. The dataset refers to individual houses advertised for which we avail information about the geographical coordinates, the requested selling price as it appears from the advertisements and a series of other house characteristics such as their size, the number of rooms and bathrooms, the typology of the building and the presence of amenities .

Figure 4 shows the geographical location of the individual points of observation and clearly displays a marked center-periphery trend of the prices with the more expansive houses concentrated in the historical area located in the city center. This empirical evidence further supports the choice of a Spatial Lag Model.

As stated in the introduction, however, this dataset suffers from the lack of a statistical sample design in the data collection process. Indeed, the dataset is just constituted by the convenience collection of the last 1,000 listing available on the real estate companies' websites without any consideration of their spatial distribution nor for the fulfillment of the requirements of a formal sample design.

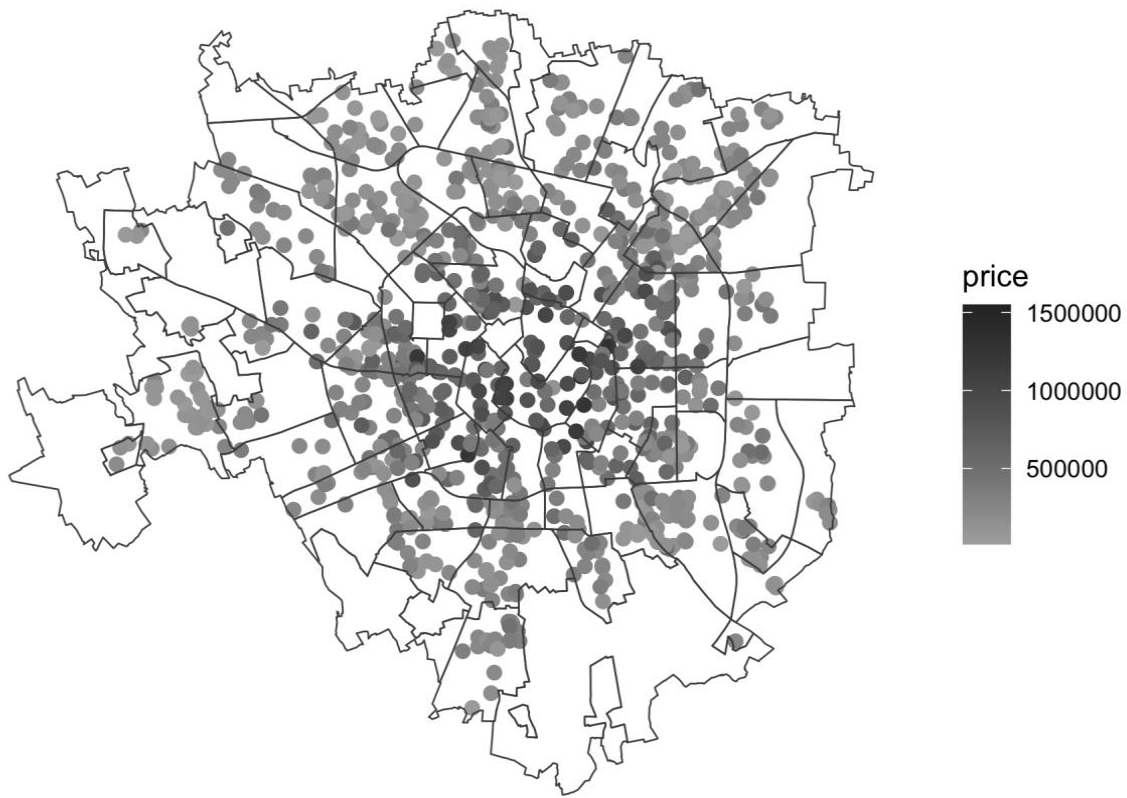

Figure 4: Map of the coordinates of the houses scraped from the real estate company websites in Milan (Italy) in May 2019. The partition refers to the 88 neighborhoods of the city of Milan (Nuclei di Identità Locali, NIL).

In order to reduce the distortions that may derive from the convenience sampling, a geographically stratified sample scheme is superimposed on the map, with the stratification units represented by the 88 neighborhoods (termed *Nuclei di Identita' Locali,* NIL, see Figure 3) and the number of families used as the stratification variable. This auxiliary information is made publicly available by Milan municipality as an open datum. Following the procedure described in the previous Section 4, a SLM is the estimated via MLE using various levels of the post-sampling parameter $\zeta = [0; 1]$.

In our exercise, we postulate a simple spatial regression model which predicts the price of a house as a function of only one independent variable (namely the size expressed in square meters) with an additional spatial lag component that is intended to capture the location effects.

The estimation results based on the procedure described in Section 4 are reported in Table 3. The table shows that the trade off between bias and estimation variance in estimating the regression slope $\beta$ is solved using the value $\zeta = 0.4$ which minimizes the MSE of the estimator. Both the estimation of $\beta$ and $\rho$ were found significant at the 5 % level.

Table 3: ML estimation of a Spatial Lag Model with varying post-sampling parameter $\zeta$.

| $\zeta$ | Sample size | $\hat{\rho}$ | $\hat{\beta}$ | Relative Bias | $MSE(\hat{\beta}, \zeta)$ |
|---|---|---|---|---|---|
| 0 | 1,000 | 0.77 | 3833.99 | 9.0% | 154687.83 |
| 0.2 | 810 | 0.73 | 3901.60 | 7.4% | 110405.05 |
| 0.4 | 610 | 0.79 | 4085.39 | 3.1% | 33022.39 |
| 0.6 | 410 | 0.83 | 3992.95 | 5.3% | 72865.23 |
| 0.8 | 250 | 0.84 | 4194.31 | 0.5% | 44804.66 |
| 1 | 220 | 0.81 | 4214.90 | 0.0% | 59591.92 |

At this MSE-optimal level of post-sampling the relative bias goes down to 3.1 % starting from the level of 9% when no post-sampling correction is considered and all 1,000 observations are included in the estimation procedure. Finally, Table 3 also displays a clear trend in the estimation of the parameter $\rho$ when we increase the level of post-sampling correction, with values that increase from 0.77 to 0.81 when $\zeta$ goes from 0 to 1.

**6. Summary, open problems and future work**

In the empirical practice, in an increasing number of cases geo-coded data, potentially very useful for estimating spatial econometric models, are collected without following any formal sampling scheme on the basis of pure convenience. Popular examples are represented by the process of crowdsourcing and web scraping. However, due to the lack of a formal sample design, in these cases parameter's estimators are of doubt reliability and, in general, lose their optimality properties (see, e. g., Hansen et al., 1953).
In particular, focusing on a Spatial Lag Model, through some Monte Carlo experiments, we have shown that the ML estimator of the regression slope is upward biased if we employ non-traditional sources and data are collected with a convenience criterion. To overcome this problem, by exploiting the general idea of *post-sampling* introduced in Arbia *et al*. (2018), we proposed to manipulate the available data prior their use in a statistical inferential context, so that they resemble a formal sample design.
Our Monte Carlo experiments on Spatial Lag Models showed that the post-sampling procedure, while having the positive effect of reducing the upward bias of the ML estimator of the regression slope, also increases the estimation variance due to the diminished degrees of freedom. An obvious solution to these contrasting results is constituted by the use of the MSE criterion so as to compromise between decreasing bias and variance increase. Employing a Maximum Likelihood approach, we showed how it is possible to monitor the increase in the estimation variance connatural to the correction mechanism at different levels of the post-sampling parameter so as to identify the optimal level of correction.
The empirical example focused on the estimation of a hedonic price model using data that are web-harvested from real estate company web advertisements in Milan (Italy). In this example we showed how is it possible to exploit the idea of post-sampling in order to reduce the bias in spatial econometric parameter estimates while moderating the loss in efficiency.
The search for solutions to the problems raised by non-probabilistic data collections (such as web-harvested or crowdsourced datasets) is, undoubtedly, still in its infancy especially with regard to their use in spatial econometrics. In this paper we only aimed at shedding light on the problem and provide some very preliminary solution being well aware that the method proposed is subject to a series of limitations that will need to be addressed in the future.

A major limitation of the post-sampling method suggested here is the fact that it is based on the assumption that we have a full knowledge of the location of all individuals at a population level (or, at least, of an auxiliary variable) so that we can use this information as the basis for stratification. This hypothesis is admittedly rarely realized in practice where the use of webdata gathering methods is exactly motivated by the lack of information about the population together with the ease of collection.

Furthermore, although in this paper we suggested an operational procedure to identify a compromise between bias and estimation variance in empirical cases, it is clear that the increase in variance which is connatural to the approach suggested in this paper represents a second major drawback in the application of the proposed method and that solutions should be devised in order to reduce it, if not eliminate it fully. One possible way out in this respect could be the use of a weighted version of the post-sampling that we discussed briefly in Section 2 whose theory, however, has not yet been developed.

Finally, the focus of the present paper has been only on the estimation of one single regression parameter $\beta$. Consequently, no consideration has been given to the effects of using convenience sampling and of the corresponding post-sampling corrections when estimating a model with more than one predictor nor the estimation of the spatial correlation parameter $\rho$. Furthermore, we also did not discuss the effects on the size and the power of the tests of hypothesis associated to the model.

All these important issues remain still open and they constitute material for further refinements of the procedure to be developed in some future work.

**Appendix: Variance of the parameter MLE estimators in a post-sampled Spatial Lag Model**

In the procedure illustrated in Section 4, in order to identify the optimal level of the post-sampling parameter $\zeta$, it is necessary to have an explicit expression for the estimation varainces. In this respect we need to derive, first of all, the Hessian associated to the likelihood function of generic element $H(\theta) = \left[\frac{\partial^2 lnL(\theta)}{\partial \theta \partial \theta^T}\right]$, L() being the model's likelihood function and $\theta$ the parameters' vector. In the case of the Spatial Lag model considered in Equation (1) we have $\theta = [\beta, \rho, \sigma^2]$. Anselin (1988) provides the exact expressions of the Hessian of a generic heteroskedastic SARAR model while Lee (2004) reports the same expressions for the homoskedastik Spatial Lag Model. From these results we have:

$H_{\beta\beta} = n\sigma^{-2} X^T X$ (A1)

$H_{\beta\rho} = n\sigma^{-2} X^T W (I - \rho W)^{-1} X\beta$ (A2)

$H_{\beta\sigma^2} = 0$ (A3)

$H_{\rho\rho} = (n^{-1} tr(W(I - \rho W)^{-1})^2 + n^{-1} tr(W(I - \rho W)^{-1})^T (W(I - \rho W)^{-1}) + n^{-1}\sigma^{-2} (W(I - \rho W)^{-1} X\beta)^T (W(I - \rho W)^{-1} X\beta)$ (A4)

$H_{\rho\sigma^2} = n^{-1}\sigma^{-2} tr(W(I - \rho W)^{-1})$ (A5)

$H_{\sigma^2\sigma^2} = \frac{1}{2} tr(\Omega^{-2}) = \frac{1}{2\sigma^4}$ (A6)

These values can be ordered in the Hessian matrix:

$$H(\theta) = \begin{bmatrix} H_{\beta\beta} & H_{\beta\varrho} & 0 \\ H_{\beta\varrho} & H_{\varrho\varrho} & H_{\varrho\sigma^2} \\ 0 & H_{\varrho\sigma^2} & H_{\sigma^2\sigma^2} \end{bmatrix} \quad (A7)$$

In principle the Fisher information matrix can be calculated as $I(\theta) = -E[H(\theta)]$ and the parameters' covariance matrix as:

$$I(\theta)^{-1} = -E[H(\theta)]^{-1} \quad (A8)$$

However, since the above expected value cannot be calculated in closed form in most practical cases because the elements of $H(\theta)$ are complicated non-linear functions of the data, we can estimate the elements of the covariance matrix in each empirical case by substituting in Equations (A1) - (A6) the vector of the ML estimators $\hat{\theta} = [\hat{\beta}, \hat{\rho}, \widehat{\sigma^2}]$ (see Greene, 2018), thus obtaining the estimated Fisher information matrix:

$$\hat{H}(\theta) = \begin{bmatrix} H_{\hat{\beta}\hat{\beta}} & H_{\hat{\beta}\hat{\rho}} & 0 \\ H_{\hat{\beta}\hat{\rho}} & H_{\hat{\varrho}\hat{\rho}} & H_{\hat{\varrho}\hat{\sigma}^2} \\ 0 & H_{\hat{\varrho}\hat{\sigma}^2} & H_{\hat{\sigma}^2\hat{\sigma}^2} \end{bmatrix} \quad (A9)$$

We can then calculate the empirical counter-part of the estimators' asymptotic covariance matrix as:

$$\hat{I}(\theta)^{-1} = -[\hat{H}(\theta)]^{-1} \quad (A10)$$

Consequently, the asymptotic variances of the various parameters' estimators, say AVar($\hat{\theta}$), are the diagonal elements of the matrix $\hat{I}(\theta)^{-1}$.

In particular, we can use Equation (A10), to calculate the asymptotic variance of $\hat{\beta}$ using the post-sampled subset of data with correction parameter $\zeta$, say $AVar(\hat{\beta})_\zeta$. This expression can then be used in Equation (3) in order to identify the optimal level of $\zeta$ which minimizes the MSE.